# An Intermediate-Mass Black Hole of ≥ 500 Solar Masses in the Galaxy ESO 243-49

Sean A. Farrell[1,2]†, Natalie A. Webb[1,2], Didier Barret[1,2], Olivier Godet[3] & Joana M. Rodrigues[1,2]

**Ultra-luminous X-ray sources are extragalactic objects located outside the nucleus of the host galaxy with bolometric luminosities[1] >$10^{39}$ erg s$^{-1}$. These extreme luminosities – if the emission is isotropic and below the theoretical (i.e. Eddington) limit, where the radiation pressure is balanced by the gravitational pressure – imply the presence of an accreting black hole with a mass of ~$10^2$–$10^5$ times that of the Sun. The existence of such intermediate mass black holes is in dispute, and though many candidates have been proposed, none are widely accepted as definitive. Here we report the detection of a variable X-ray source with a maximum 0.2–10 keV luminosity of up to 1.2 x $10^{42}$ erg s$^{-1}$ in the edge-on spiral galaxy ESO 243-49, with an implied conservative lower limit of the mass of the black hole of ~500 $M_{sun}$. This finding presents the strongest observational evidence to date for the existence of intermediate mass black holes, providing the long sought after missing link between the stellar mass and super-massive black hole populations.**

Apart from the super-Eddington luminosities of ultra-luminous X-ray (ULX) sources, other circumstantial evidence exists for intermediate mass black holes. Low-temperature disc blackbody spectral components and quasi-periodic oscillations (QPOs) believed to originate at the inner regions of the accretion disc are seen in some ULXs, implying black hole masses above the stellar range[2,3]. However, neither can be conclusively linked to the innermost disc radius, precluding a definitive mass estimate. In the globular cluster ω Cen the presence of an intermediate mass black hole was derived from the rise in the velocity dispersion profile towards the cluster centre, although a concentration of stellar remnants or a radially biased orbital structure are also valid explanations[4]. Modelling showed that the cluster dynamics and the current mass are consistent with an accreted stripped dwarf galaxy nucleus[5] with a mass of ~$10^7$ $M_{sun}$. The blackbody X-ray spectrum and bolometric luminosity of ~$10^{39}$ erg s$^{-1}$ of the ultra-luminous super-soft-source in M81 may imply the presence of an intermediate mass black hole[6], but the source properties are also consistent with a massive white dwarf accreting at a high rate. The luminosity can alternatively be explained through mild super-Eddington accretion and/or beaming[7]. In addition, the black hole mass was derived by estimating the shape of the X-ray spectrum from widely spaced optical observations. This assumes a smooth continuation between the optical and X-ray spectra, and that the spectra were stable over the ~6 year time-span of the observations. The latter assumption is inconsistent with observational data, as black holes of all sizes demonstrate significant spectral variability over a wide range of timescales[8].

Whilst investigating the 2XMM serendipitous source catalogue[9], we identified 2XMM J011028.1-460421 (HLX-1 for simplicity) at right ascension 01h 10min 28.2s, declination –46° 04' 22.2" (J2000) with a 3σ positional uncertainty of 0.83'', observed on 2004 November 23 as part of Obs-Id 0204540201. The position of HLX-1 is coincident with the absorption line galaxy ESO 243-49, at a redshift[10] of $z$ = 0.0224 ± 0.0001, 8" from the nucleus yet within the confines of the galaxy (Figure 1). The position and error in each observation were derived in accordance with the methods used for the creation of the 2XMM catalogue[9], with the final position and error determined by taking the weighted mean of the two positions. No indication for disruption of the galactic disc, as expected following a galaxy merger, is visible. Contrary to ω Cen, HLX-1 is unlikely to represent the stripped nucleus of an accreted dwarf galaxy. The XMM-Newton European Photon Imaging Camera spectra (Figure 2) of HLX-1 are best fitted by an absorbed power-law model, with parameters consistent with other ULXs[1] (Table 1). The derived luminosity (assuming the ESO 243-49 distance) is ~400 times larger than the Eddington value for a 20 $M_{sun}$ black hole, and almost an order of magnitude greater than that of the previously reported most luminous ULX[11] (0.2–10 keV unabsorbed $L_x < 2 \times 10^{41}$ erg s$^{-1}$). A deeper follow-up observation performed with XMM-Newton on 2008 November 28 (Obs-Id: 0560180901) showed a change in the luminosity and the spectral shape (Figure 2). Fourier power spectra were computed using the pn light curves for this observation (6 ms resolution) to test for high-frequency QPOs. No significant modulation was found at frequencies <83 Hz. Fitting of the spectrum required the addition of a soft disc blackbody component ($\Delta\chi^2 = 155$ for 2 degrees of freedom less, Table 1), consistent with other ULXs. Adding this component to the fit of observation 1 did not improve the $\chi^2$ significantly ($\Delta\chi^2 = 2.1$ for 2 degrees of freedom less).

[1]Université de Toulouse, UPS, CESR, 9 Avenue du Colonel Roche, F-31028 Toulouse Cedex 9, France. [2]CNRS, UMR5187, F-31028 Toulouse, France. [3]Department of Physics and Astronomy, University of Leicester, University Road, Leicester, LE1 7RH, UK. †Present address: Department of Physics and Astronomy, University of Leicester, University Road, Leicester, LE1 7RH, UK.

Using the posterior predictive p-values method[12], we simulated 5,000 data sets with exposure times, background, and energy response identical to the first observation using the best fit spectral parameters of the second observation, and fitted them with an absorbed power law. In less than 1% of the case the power law fit is acceptable and the fitted parameters are consistent with the values reported in Table 1 for the first observation (within the 90% confidence errors). Further, using the same method and an additional 5,000 simulations, we evaluated the significance of the disc blackbody component in the first observation to be ~70%, insufficient to claim it is real. These simulations demonstrate unambiguously that HLX-1 has varied between the two observations.

The case for the hyper-luminous nature of HLX-1 depends on its association with ESO 243-49. The probability that any of the real 2XMM sources surrounding ESO 243-49 could be randomly aligned with any of the background galaxies in the field, calculated by 1,000,000 Monte Carlo simulations, is ~9%. However, by taking into account the X-ray spectral shape we can confidently rule out a foreground source. The X-ray spectrum lets us exclude coronal emission from a star, which is typically fitted by multi-temperature thermal models[13]. Non-thermal emission has been detected from supernovae and from shock-boundaries in shell-type supernova remnants[14] with X-ray luminosities between $10^{37}$–$10^{41}$ erg s$^{-1}$. However, the non-thermal flux decays rapidly after the supernova explosion[14], with thermal emission dominating after ~100 days. A Galactic white dwarf accreting from a low mass companion could be visible in X-rays but not show up in our optical image. The spectra of these objects are consistent with emission from a thermal plasma with temperatures >2 keV and non-zero elemental abundances[15,16]. The spectra from both observations are inconsistent with this model. A quiescent neutron star low mass X-ray binary is also inconsistent with our data, as the spectrum should be thermal (representing emission from the neutron star atmosphere) possibly with the addition of a flat power law component[17]. The steep power-law spectrum of HLX-1 during observation 1 is not consistent with a neutron star in either the high or low states, and instead indicates the presence of an accreting black hole in the high/soft state[18]. The derived luminosity for HLX-1 at 10 kpc is only ~$10^{34}$ erg s$^{-1}$, too low for a Galactic black hole binary undergoing near-Eddington accretion. Therefore HLX-1 cannot be a foreground Galactic object.

Some blazars (a sub-class of radio-loud active galactic nuclei with the jet pointing close to our line of sight) have steep power-law spectra[19]. However, even the faintest blazars have flux densities >2 mJy at 1.4 GHz (ref.20). Previous radio observations[10] detected a source consistent with the core of ESO 243-49 with a 1.4 GHz flux density of 0.16 ± 0.03 mJy, inconsistent with the position of HLX-1 (Figure 1). The relatively low intrinsic absorption also rules-out observing a blazar through the disc of ESO 243-49, yet is consistent with a number of other ULXs[21]. The association of HLX-1 with ESO 243-49 is almost certain, and thus our derived luminosity valid.

The observed variability between observations indicates that HLX-1 cannot be a superposition of multiple lower-luminosity sources. The maximum luminosity derived from the first observation therefore implies a mass lower limit of ~5,400 $M_{sun}$, assuming Eddington accretion and isotropic emission. Relativistic jets have been observed in a number of Galactic black hole X-ray binaries during the low/hard state[8], where the accretion rate and luminosity decrease and the spectrum flattens. If the source is viewed at angles close to the jet axis, relativistic boosting could amplify the flux so that the luminosity appears to exceed the Eddington value. Modelling of stellar mass black hole binaries has shown that relativistic boosting could produce apparent luminosities five times the Eddington limit[22], implying a black hole mass of >1,000 $M_{sun}$ for HLX-1. However, relativistic boosting is unlikely as a flat spectrum – as opposed to the steep spectrum we observe in the first observation – is predicted[22], and still requires an intermediate mass black hole for luminosities >$10^{41}$ erg s$^{-1}$.

Geometric beaming whereby a thick accretion disc collimates the X-ray emission coupled with super-Eddington accretion can explain apparent luminosities up to ~$10^{40}$ erg s$^{-1}$, with higher luminosities for hydrogen-poor accretion[7]. A 20 $M_{sun}$ black hole with an apparent luminosity of $10^{42}$ erg s$^{-1}$ implies a low beaming factor of $b \approx 0.01$, with a mass accretion rate of ~2.6 times the Eddington value for hydrogen-poor accretion[7]. However, the derived luminosity is high enough above the Eddington limit that matter would be easily accelerated to Lorentz factors of ~5–10, leading to relativistic jets and flat spectra[23]. The combination of the derived luminosity and steep spectrum of HLX-1 in the first observation thus rules out relativistic and geometric beaming. Accreting stellar mass black holes may exceed the Eddington luminosity by up to a factor of ten[24], implying an Eddington luminosity of ~$10^{41}$ ergs s$^{-1}$ (0.2–10 keV), and a lower mass limit of ~500 $M_{sun}$ for HLX-1. This limit is derived assuming a bolometric luminosity, while our derived luminosity is in the 0.2–10 keV band. The bolometric luminosity is likely to exceed our value, so our mass lower limit is a conservative estimate. HLX-1 therefore presents the strongest case to date for the existence of intermediate mass black holes.

**Acknowledgements**
We thank N. Schartel for granting an observation under the XMM-Newton project scientist discretionary time program. We thank R. Belmont, A. King, J-P. Lasota, K. Mukai, T. Roberts, S. Rosen, S. Sembay and M. Watson for useful discussions. S.A.F acknowledges funding from the CNES. S.A.F. and O.G. acknowledge STFC funding. This work made use of the 2XMM Serendipitous Source Catalogue constructed by the XMM-Newton Survey Science Centre on behalf of ESA. We thank the Swift team for performing a TOO observation which provided justification for an additional observation with XMM-Newton. This work was based on observations obtained with XMM-Newton, an ESA science mission with instruments and contributions directly funded by ESA Member States and NASA.

**Author Information**
Reprints and permissions information is available at www.nature.com/reprints. The authors declare no competing financial interests. Correspondence and requests for materials should be addressed to S.A.F. (saf28@star.le.ac.uk) or N.A.W (natalie.webb@cesr.fr).

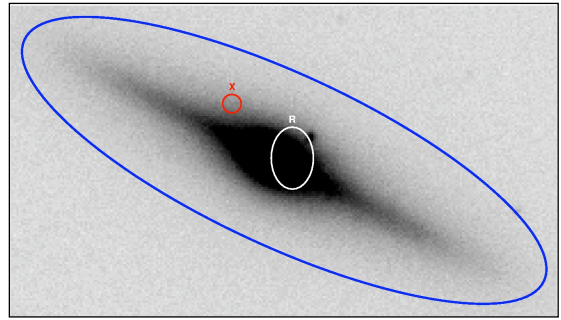

**Figure 1 | R-band optical image of ESO 243-49 obtained with the Very Large Telescope.** The centre of the red circle (X) indicates the position of HLX-1, with the radius representing a 3σ positional uncertainty of 0.83". The centre of the white ellipse (R) indicates the position of the radio detection from the Phoenix Deep Survey[10], with the radii representing the 3σ uncertainty. The blue ellipse indicates the elliptical confines of the galaxy.

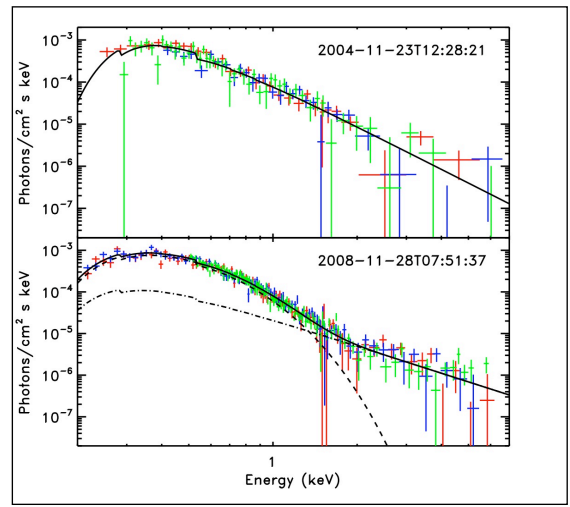

**Figure 2 | European Photon Imaging Camera X-ray spectra of HLX-1.** Top panel: the unfolded pn (green), MOS1 (red) and MOS2 (blue) spectra of the first observation fitted with the best-fit absorbed power-law model (solid black line). Bottom panel: the unfolded spectra of the second observation, fitted with the best-fit absorbed power law (black dot-dashed line) plus disc blackbody (dashed line) model. The error bars indicate the 90% confidence limits.

**Table 1 | X-ray spectral parameters of HLX-1**

| Parameter | Observation 1 | | Observation 2 | | Unit |
|---|---|---|---|---|---|
| $nH_{total}$ | 0.08 | +0.03 / −0.03 | 0.04 | +0.01 / −0.01 | $10^{22}$ atoms cm$^{-2}$ |
| $nH_{intrinsic}$ | 0.06 | +0.03 / −0.03 | 0.02 | +0.01 / −0.01 | $10^{22}$ atoms cm$^{-2}$ |
| $T_{in}$ | … | | 0.18 | +0.01 / −0.01 | keV |
| $Norm_{DBB}$ | … | | 29 | +6 / −5 | |
| $\Gamma$ | 3.4 | +0.3 / −0.3 | 2.2 | +0.4 / −0.3 | |
| $Norm_{PL}$ | 9 | +1 / −1 | 2.2 | +0.8 / −0.6 | $10^{-5}$ photons cm$^{-2}$ s$^{-1}$ keV$^{-1}$ |
| $\chi^2/dof$ | 113.4/108 | | 333.9/329 | | |
| $L_x$ | 11 | +0.1 / −4.0 | 6.4 | +0.6 / −0.6 | $10^{41}$ erg s$^{-1}$ |

$nH_{total}$ = total neutral hydrogen column density; $nH_{intrinsic}$ = neutral hydrogen column density intrinsic to the source, after taking into account the Galactic absorption[25] of 0.0179 x $10^{22}$ atoms cm$^{-2}$; $T_{in}$ = temperature at inner disc radius; $Norm_{DBB}$ = normalisation of disc blackbody component; $\Gamma$ = power-law photon index; $Norm_{PL}$ = normalisation of power law component; $\chi^2/dof$ = $\chi^2$ statistic and associated degrees of freedom (*dof*) for the spectral fitting; $L_x$ = unabsorbed X-ray luminosity in the 0.2–10 keV band.